\documentclass[prb, 12pt, preprint]{revtex4-1}
\usepackage{graphicx}
\begin{document}
\title{Upper Limit Transverse Voltage Calculations from the Glazman Model.} 
\author{M.E. Broussard, and P.R. Broussard}
\email{phill.broussard@covenant.edu}
\affiliation{Covenant College, Lookout Mountain, GA 30750}

\begin{abstract}

The model of Transverse Voltage proposed by L.I. Glazman made predictions on the upper current limit for the existence of Transverse Voltage.  In his paper, the upper limit for current at which Transverse Voltage would appear in wide films is temperature independent. Using a Runge-Kutta algorithim and the complete equations for vortex-antivortex interactions,  the upper limit for thin niobium films were calculated for several temperatures, and how these compare to Glazman's model and experimental results on the same films are discussed.  We find in contrast to both the experimental results and Glazman's prediction, the solution to the complete equations show an increase in the upper current limit as temperature increases.
\end{abstract}

\maketitle

\section{Introduction}

Glazman's model\cite{Glazman} on Transverse Voltage (TV), observed in thin superconducting films near their transition temperature, considered the annihilation of vortices and antivortices as the cause of TV.  He used a simplified model for the equations of motion of a vortex-antivortex (VA) pair in a superconducting film, and came up with an expression for the maximum current at which TV would occur.  In this paper, we numerically solve the complete equations for a VA pair in a wide film and calculate the maximum current.  We will compare our results to what Glazman predicted as well as see how both predictions compare to experimental results on TV measurements in two niobium films.\cite{Spencer}

\section{Glazman model}

Glazman considered two cases in his original paper,\cite{Glazman} the case of narrow films ($w\ll2\lambda^{2}/d$) and wide films ($w\gg2\lambda^{2}/d$), where $w$ is the film width, $\lambda$ is the London penetration depth for the material, and $d$ is the film thickness.  Since the experimental data that we will compare to is for the wide film case, we will focus on this part of Glazman's paper.  In his model, the TV signal begins at a current $I_{c1}$ when vortices can enter a film, which is given by Likharev\cite{Likharev} for the wide film case as
\begin{equation}
\label{Lik}
I_{c1}=\frac{\Phi_0d}{2\pi\mu_0\lambda_0^2}(1-t^4)\ln2\frac{\lambda_0^2}{d\xi_0}\frac{\sqrt{1-t}}{(1-t^4)},
\end{equation}
where $\Phi_{0}$ is the flux constant $h/2e$, $\xi_0$ is the zero temperature Ginzburg Landau coherence length, $\lambda_0$ is the zero temperature London penetration length, $t$ is the reduced temperature ($T/T_c$), and we are using the MKS system of units.

Glazman then derived an expression for the maximum current where TV would appear, given by $I_{c1}(\frac{\ell_{0}}{\ell})^{2}$, where $\ell$ is the longitudinal distance between vortex and anti-vortex when they enter the film and $\ell_{0}$ is the maximum length the vortex and anti-vortex could be separated at the start and still combine. By setting the Lorentz driving force and VA driving force equal to each other, Glazman derived an equation for $\ell_{0}$, which when combined with $I_{c1}$ gives the maximum current as $\frac{\Phi_{0} w}{\pi \mu_{0} \ell^2}$. This results in the maximum current being temperature independent.

In deriving the above result, Glazman used an approximation to the VA attractive force.  We desired to calculate the upper limit of the current by using the general form of the VA attraction\cite{Orlando} and numerically solving the equations to see at what current is a collision between the vortex and anti-vortex no longer possible.  The equations of motion for the relative coordinate of the vortex and anti-vortex pair would be given by (where these are equations of force, not force per unit length as Glazman used)
\begin{eqnarray}
\eta\dot{x}/2 & = &-\frac{\Phi^2_0 d}{2\pi\mu_0\lambda^3}K_1(\frac{\sqrt{x^2 + y^2}}{\lambda})(\frac{x}{\sqrt{x^2 + y^2}})\\
\eta\dot{y}/2 & = &-\frac{\Phi^2_0 d}{2\pi\mu_0\lambda^3}K_1(\frac{\sqrt{x^2 + y^2}}{\lambda})(\frac{y}{\sqrt{x^2 + y^2}}) - \frac{\Phi_0 I}{w}
\end{eqnarray}
where $\eta$ is the viscosity coefficient of an individual vortex in a superconducting film, given by $\frac{\Phi_{0}^2 d}{2\pi \xi^2 \rho_0}$,\cite{Orlando} $\rho_0$ is the resistivity of the film at low temperatures, $I$ is the current passing through the film, $x$ is the longitudinal separation distance, $y$ is the transverse separation distance, and $K_{1}$ is the modified Bessel function of order 1.  Current is assumed to flow purely in the longitudinal $x$ direction and variations in current flow direction were not accounted for.  Even though that is a limitation for this model, with the VA interaction range being of the order of $\lambda$ and interactions happening near the center of the film, the current will be well approximated there as flowing in one direction as the above equations assume.

\begin{figure}
\includegraphics[width=6in]{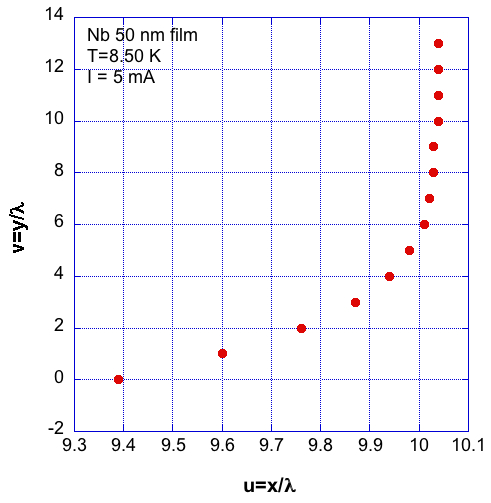}
\caption{\label{uv} Results of finding the VA collision parameters for a 50 nm Nb film at a temperature of 8.50 K with 5 mA flowing.  The graph shows the maximum allowed values of $u=x/\lambda$ and $v=y/\lambda$ where a collision between a VA pair would collide.}
\end{figure}

We then transform the above equations into dimensionless variables, with $u = x/\lambda, v = y/\lambda,  \tau = t(\frac{2\rho_0 \xi^2}{\lambda^4 \mu_0}),$
to produce the following dimensionless equations of motion:
\begin{eqnarray}
\frac{du}{d\tau} & = &-K_1(\sqrt{u^2 + v^2})(\frac{u}{\sqrt{u^2 + v^2}})\\
\frac{dv}{d\tau} & = &-K_1(\sqrt{u^2 + v^2})(\frac{v}{\sqrt{u^2 + v^2}}) - \frac{2\pi I \mu_0 \lambda^3}{\Phi_0 w d}
\end{eqnarray}
We cannot use the same graphical analysis Glazman used in his paper to find the colliding vs. non-colliding trajectories, and instead must use a Runge-Kutta algorithm to solve the above equations.   This numerical approach can determine under what conditions the vortices would collide (decided by the separation distance $\le 0.01 \lambda$) by inputting the dimensions and properties, the current being passed through and the temperature of the thin film. Then using the data generated by the software we are able to predict the upper limit of the starting $u$ value when a collision would occur for a given $v$ value.  In Fig.\ref{uv} we show the result of running the program for a 50 nm niobium film at a temperature of 8.50 K with 5 mA current flowing, and assuming $\lambda_0=26$ nm.  For this set of parameters, we can find the maximum $u$ value, or maximum separation between vortex and anti-vortex for which a collision is possible, which for this case is $u_{\mathrm{max}} \approx 10.05$.  Such curves were produced for the above mentioned 50 nm Nb film and a 25 nm Nb film, whose measured TV data is given in the work by Spencer and Broussard.\cite{Spencer}

\begin{figure}
\includegraphics[width=6in]{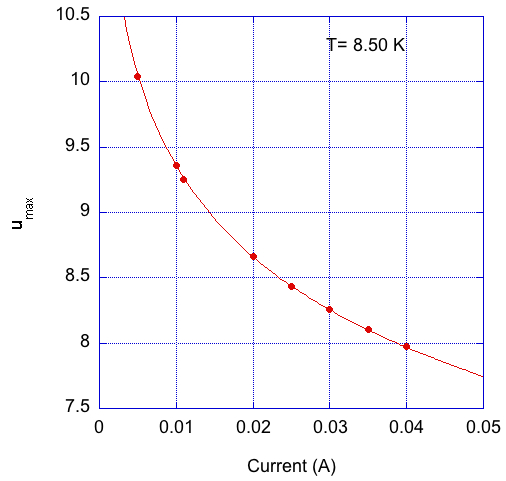}
\caption{\label{xT} The maximum allowed horizontal separation, $u_{\mathrm{max}}$, vs current for 50 nm Nb film at 8.50 K.  The curve is a fit to the curve $\ln(I_0/I)$. For this data set, $I_0$ is 115.6$\pm$0.2 Amps and $\chi^{2}$ is less than 0.0001.}
\end{figure}

A plot such as that in Fig.\ref{uv} would then be generated for 8 different currents for each temperature.  In Fig.\ref{xT} we show the results for a temperature of 8.50 K for the 50 nm Nb film, giving the maximum longitudinal separation between the vortex and anti-vortex allowed for that current.  We find we can fit this curve, as well as the others generated for three other temperatures as well as the results for the 25 nm Nb film, to a form of $u_{\mathrm{max}}= \ln(I_{0}/I),$ with $\chi^{2}$ values being $\leq0.007$ for both films and all temperatures.   

\begin{figure}
\includegraphics[width=6in]{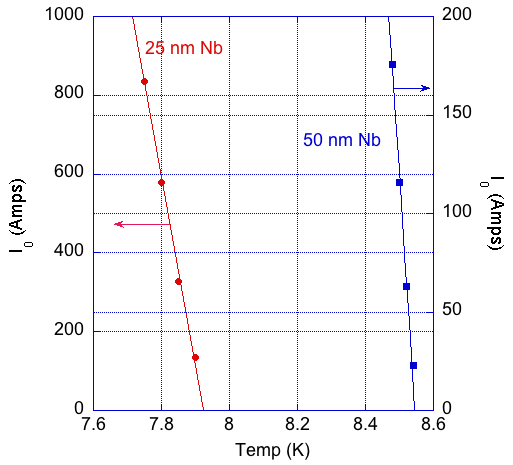}
\caption{\label{IT} Plot of $I_0$ values for the 25 nm  and 50 nm Nb films vs temperature, with linear fits. }
\end{figure}

We can take the computed $I_0$ values from the fits and plot them as a function of the temperature of the two films and see what behavior is observed.  This is done in Fig.\ref{IT}, where a clear linear dependence is observed.  From these we can determine the $T_c$ of the films, which is found to be 7.93 K for the 25 nm film and 8.55 K for the 50 nm film.  These compare very well with the experimentally determined $T_c$ found from the midpoint of the resistive transition, which were 7.91 K for the 25 nm films and 8.54 K for the 50 nm film.  That we are seeing a consistent trend in the fits is an encouraging sign, however the large values of the derived currents is strange.  We could not come up with a simple understanding for why the values of $u_{\mathrm{max}}$ fit such an equation.

\begin{figure}
\includegraphics[width=6in]{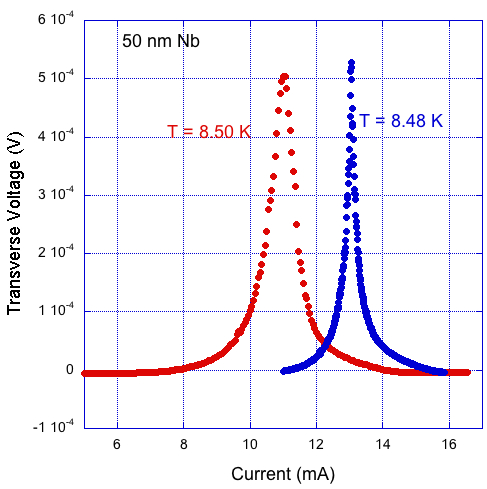}
\caption{\label{Exp1} Experimental TV signals from a 50 nm Nb sample at two temperatures.\cite{Spencer} }
\end{figure}

We then assumed as Glazman assumed that there is a fixed longitudinal separation, $\ell$, between the VA pair when they enter the film.  So when the value of $x_{\mathrm{max}}=\lambda u_{\mathrm{max}}>\ell$ then there should no longer be a TV signal.  Experimentally, the TV signals do not go suddenly to zero, but instead gradually approach zero, as shown in Fig.\ref{Exp1} for the 50 nm Nb film.  So to get an experimental measure of the upper limit of the TV signal, we chose to define the maximum current when the TV signal had fallen to 3\% of the maximum signal.  For the four temperatures for each Nb film, we can then define a current which we denote as the upper limit of TV signal.  As seen in Fig.\ref{Exp1} this current decreases as the temperature increases.  From Glazman's paper, this upper limit should be temperature independent, which clearly does not agree with the experimental results.  For our model with the complete VA interaction, what dependence is seen?  

\begin{figure}
\includegraphics[width=6in]{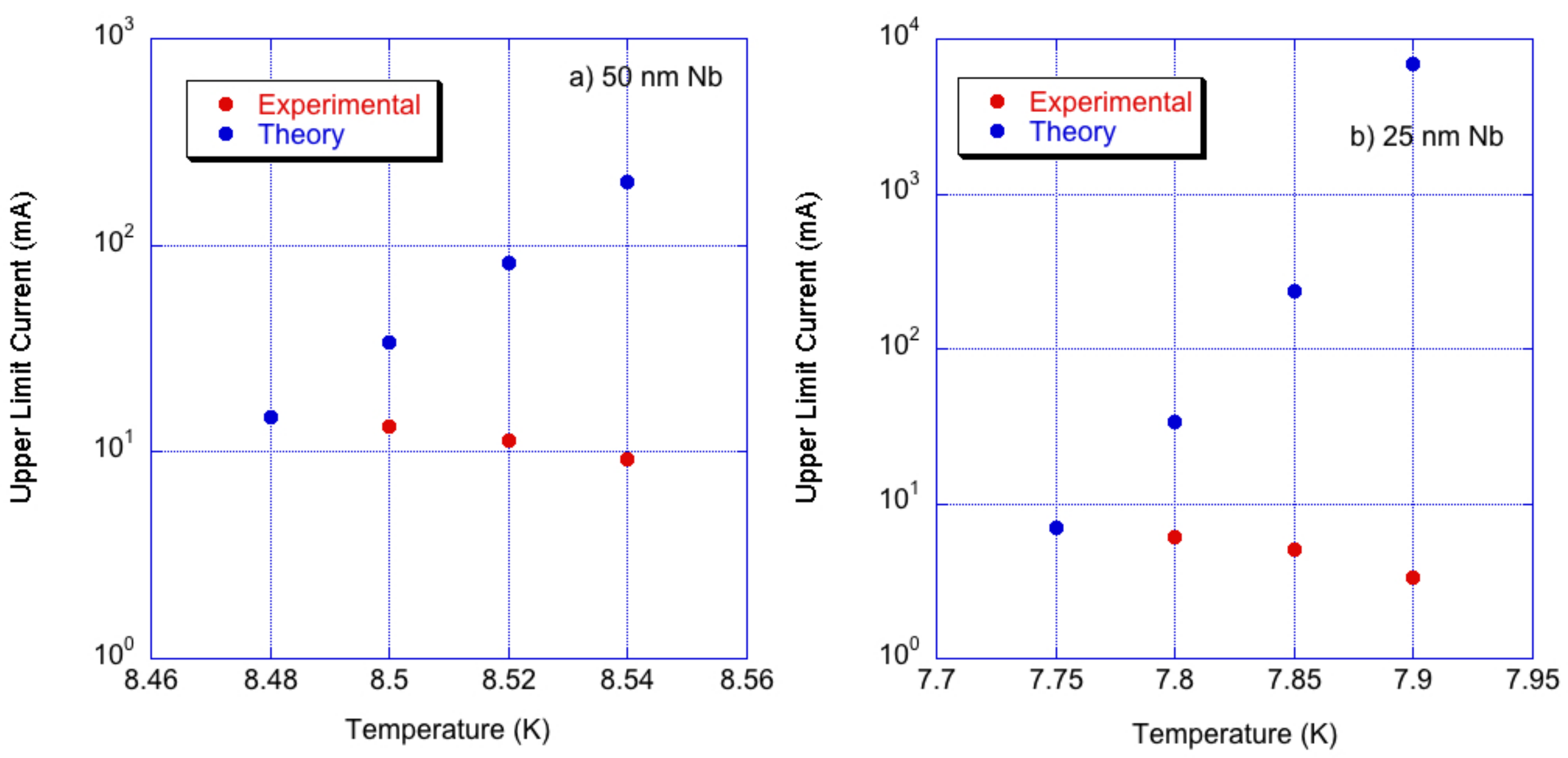}
\caption{\label{IvsT} Comparison of maximum current for TV for a) a 50 nm Nb film and b) a 25 nm Nb film.  The y-axis is in a Log scale.  In both plots we show the experimental maximum current values (labeled Experiment) and those determined by our model (labeled Theory).  The prediction from Glazman would be a constant value for all temperatures.}
\end{figure}

For our model, we use the experimental value of maximum current at the lowest temperature for each Nb film and the corresponding plot like Fig.\ref{xT} to give us a corresponding value of $u_{\mathrm{max}}$.  We then scale that value with the value of $\lambda$ at that temperature to find the resulting value of $x_{\mathrm{max}}$ that corresponds to this temperature and will be the film's value of $\ell$, the VA longitudinal separation when they enter the film.  For the 50 nm Nb film, for example, the value of $\ell$ is about 1.46 $\mu$m and for the 25 nm sample it is about 0.92 $\mu$m. Assuming this value of $\ell$ stays constant, we can find the corresponding  new values of $u_{\mathrm{max}}$ at higher temperatures by scaling with the corresponding values of $\lambda$ at those temperatures, and from Fig. \ref{xT} find the current that correspond to that new value of $u_{\mathrm{max}}$.  What is seen, as shown in Fig. \ref{IvsT} is that the predicted maximum current for which TV is possible by VA collision {\bf increases} as temperature goes up, while Glazman predicted it would be temperature independent, and experimentally we find the maximum current decreases as the temperature increases.   This result would not be limited to just these two Nb films presented in this paper.  Since the basic physics is set by the details of the VA interaction, as well as the temperature dependence of $\lambda$ and $\xi$, it is clear that this will be the case for any film described by these equations.  

\section{Conclusion}

As seen in our study, using the full VA interaction and Glazman's approximation of a fixed separation between the VA pair for the wide film case results in upper limit currents for TV increasing as the temperature increases, in contradiction to Glazman's model that such currents would be temperature independent, as well as experimental results which show they decrease as the temperature increases.  Without a strong temperature dependence to the value of VA separation, it is hard to see how the Glazman model can model the experimental data observed in TV studies.  So at this point, the explanation by Glazman does not seem to correspond to experimental results.

\end{document}